\documentclass[letterpaper]{article}
\usepackage{aaai}
\usepackage{times}
\usepackage{helvet}
\usepackage{courier}
\usepackage{tikz}
\usetikzlibrary{calc}
\usetikzlibrary{backgrounds,positioning}
\usetikzlibrary{decorations.pathreplacing}
\usepackage{subcaption}
\usepackage{diagbox}
\usepackage{comment}

\usepackage{xcolor}
\usepackage[linesnumbered,ruled,vlined,procnumbered]{algorithm2e}
\usepackage{etoolbox}

\makeatletter
\AtBeginEnvironment{procedure}{\let\c@algocf\c@procedure}
\makeatother
\usepackage{amsmath}

\DeclareMathOperator*{\argmin}{argmin}

\SetCommentSty{mycommfont}
\begin{document}
%
\title{A Particle Swarm Based Algorithm for Functional \\Distributed Constraint Optimization Problems}
\author{Moumita Choudhury, Saaduddin Mahmud \normalfont and \bf Md. Mosaddek Khan\\
Department of Computer Science and Engineering, University of Dhaka\\
\{moumitach22, saadmahmud14\}@gmail.com, mosaddek@du.ac.bd}
\maketitle
\begin{abstract}
Distributed Constraint Optimization Problems (DCOPs) are a widely studied constraint handling framework. The objective of a DCOP algorithm is to optimize a global objective function that can be described as the aggregation of a number of distributed constraint cost functions. In a DCOP, each of these functions is defined by a set of discrete variables. However, in many applications, such as target tracking or sleep scheduling in sensor networks, continuous valued variables are more suited than the discrete ones. Considering this, Functional DCOPs (F-DCOPs) have been proposed that is able to explicitly model a problem containing continuous variables. Nevertheless, the state-of-the-art F-DCOPs approaches experience onerous memory or computation overhead. To address this issue, we propose a new F-DCOP algorithm, namely Particle Swarm Based F-DCOP (PFD), which is inspired by a meta-heuristic, Particle Swarm Optimization (PSO). Although it has been successfully applied to many continuous optimization problems, the potential of PSO has not been utilized in F-DCOPs. To be exact, PFD devises a distributed method of solution construction while significantly reducing the computation and memory requirements. Moreover, we theoretically prove that PFD is an anytime algorithm. Finally, our empirical results indicate that PFD outperforms the state-of-the-art approaches in terms of solution quality and computation overhead.
\end{abstract}
\section{Introduction}
Distributed Constraint Optimization Problems (DCOPs) are an important constraint handling framework of multi-agent systems in which multiple agents communicate with each other in order to optimize a global objective. The global objective is defined as the aggregation of cost functions (i.e. constraints) among the agents. The cost functions can be defined by a set of variables controlled by the corresponding agents. DCOPs have been widely applied to solve a number of multi-agent coordination problems including, multi-agent task scheduling \cite{sultanik2007modeling}, sensor networks \cite{farinelli2014agent}, multirobot coordination \cite{Yedidsion2016ApplyingDT}.\par
Over the years, a number of algorithms have been proposed to solve DCOPs which includes both exact and non-exact algorithms. Exact algorithms, such as ADOPT \cite{modi2005adopt}, DPOP \cite{Petcu2005ASM} and PT-FB \cite{litov2017forward}, are designed in such a way that provide the global optimal solutions of a given DCOP. However, exact solutions experience either, or both, of the exponential memory requirements and computational cost as the system grows. On the contrary, non-exact algorithms such as, DSA \cite{zhang2005distributed}, MGM \& MGM2 \cite{Maheswaran2004Distributed}, Max-Sum \cite{farinelli2008decentralised}, CoCoA \cite{Leeuwen2017CoCoAAN}, and ACO\_DCOP \cite{chen2018ant} compromise some solution quality for scalability.\par

In general, DCOPs assume that participating agents' variables are discrete. Nevertheless, many real world applications (e.g. target tracking sensor orientation \cite{fitzpatrick2003distributed}, sleep scheduling of wireless sensors \cite{hsin2004network}) can be best modelled with continuous variables. Therefore, for discrete DCOPs to be able to apply in such problems, we need to discretize the continuous domains of the variables. However, the discretization process needs to be coarse for a problem to be tractable and must be sufficiently fine to find high quality solutions of the problem\cite{stranders2009decentralised}. To overcome this issue, Stranders et al. has proposed a continuous version of DCOP which is later referred as Functional DCOP (F-DCOP) \cite{hoang2019new}.
There are two main differences between F-DCOP and DCOP. Firstly, instead of having discrete decision variables, F-DCOP has continuous  variables that can take any value between a range.  Secondly, the constraint functions are represented in functional forms in F-DCOP rather than in the tabular forms in DCOP.\par

To cope with the modification of the DCOP formulation, Continuous Max-Sum (CMS) has been proposed which is an extension of the discrete Max-Sum \cite{stranders2009decentralised}. However, this paper approximates the constraint utility functions as piece-wise linear functions which is often not applicable in practice since a handful of real life applications deals with only peice-wise linear functions.
To address this limiting assumption of CMS, Hybrid Max-Sum (HCMS) has been proposed in which continuous non-linear optimization methods are combined with the discrete Max Sum algorithm \cite{voice2010hybrid}. However, continuous optimization methods such as, gradient based optimization require derivative calculations, and thus they are not suitable for non differentiable optimization problems.
The latest contribution in this field has been done by Hoang et al., 2019. In this paper, authors propose one exact, Exact Functional DPOP (EF-DPOP) and two approximate versions, Approximate Functional DPOP (AF-DPOP), and Clustered AF-DPOP (CAF-DPOP) of DPOP for solving F-DCOP \cite{hoang2019new}. The key limitation of these algorithms is that both AF-DPOP and CAF-DPOP incur exponential memory and computation overhead even though the latter cuts the communication cost by providing a bound on message size.\par

Against this background, we propose a Particle Swarm Optimization based F-DCOP algorithm(PFD). Particle Swarm Optimization (PSO) \cite{eberhart1995particle} is a stochastic optimization technique inspired by the social metaphor of bird flocking that has been successfully applied to many optimization problems such as Function Minimization \cite{shi1999empirical}, Neural Network Training \cite{zhang2007hybrid} and Power-System Stabilizers Design Problems \cite{abido2002optimal}. However, to the best of our knowledge no previous work has been done to incorporate PSO in DCOP or F-DCOP. In PFD, agents cooperatively keep a set of particles where each particle represents a candidate solution and iteratively improve the solutions over time. Since PSO requires only primitive mathematical operators such as, addition and multiplication, it is computationally inexpensive (both in memory and speed) than the gradient based optimization methods. Specifically, We empirically show that PFD can not only find better solution quality by exploring a large search space but it is also computationally inexpensive both in terms of memory and computation cost than the existing FDCOP solvers.
\section{Background and Problem Formulation}
In this section, we formulate the problem and discuss the background necessary to understand our proposed method. We first describe the general DCOP framework and then move on the F-DCOP framework which is the main problem that we want to solve. We then discuss the centralized PSO algorithm and the challenges remain to incorporate PSO with F-DCOP framework. 
\subsection{Distributed Constraint Optimization Problem}
A DCOP can be defined as a tuple $ \langle A,X,D,F,\alpha \rangle $ \cite{modi2005adopt} where,
\begin{itemize}
    \item A is a set of agents $\{a_1,a_2,...,a_n\}$.
    \item X is a set of discrete variables $\{x_1,x_2,...,x_m\}$, where each variable $x_i$ is controlled by one of the agents  $a_i$ $\in$ $A$.
    \item D is a set of discrete domains $\{D_1, D_2,...,D_m\}$, where each $D_i$ corresponds to the domain of variable $x_i$.
    \item F is a set of cost functions $\{f_1,f_2,...,f_l\}$, where each $f_i \in F$ is defined over a subset $x^i$ = \{$x_{i_{1}}$, $x_{i_{2}}$, ..., $x_{i_{k}}$\} of variables X and the cost for the function $f_i$ is defined for every possible value assignment of  $x^i$, that is, $f_i$: $D_{i_{1}}$ $\times$ $D_{i_{2}}$ $\times...\times$ $D_{i_{k}}$ $ \to \!R$. The cost functions can be of any arity but for simplicity we assume binary constraints throughout the paper.
    \item $\alpha: X \rightarrow
	A$ is a variable to agent mapping function which assigns the control of each variable $x_i \in X$ to an agent $a_i$ $\in$ $A$. Each agent can hold several variables. However,  for the ease of understanding, in this paper we assume each agent controls only one variable.
\end{itemize}
The solution of a DCOP is an assignment $X^*$ that minimizes the sum of cost functions as shown in Equation \ref{eq:1}. 
\begin{equation}
    X^* = \argmin_X \sum_{i=1}^{l} f_i(x^i)
\label{eq:1}
\end{equation}

\subsection{Functional Distributed Constraint Optimization Problem}
\begin{figure}

  \begin{tikzpicture}
        [
        roundnode/.style={circle, draw=green!60, fill=green!5, very thick, minimum size=7mm},
        ]
        \node[roundnode]    at(1,0)  (x1)                              {$x_1$};
        \node[roundnode]    at(0,-1)  (x2)                              {$x_2$};
        \node[roundnode]    at(1,-1)  (x3)                              {$x_3$};
        \node[roundnode]    at(2,-1)  (x4)                              {$x_4$};
         
        \draw (x1) -- (x2);
        \draw (x1) -- (x3);
        \draw (x3) -- (x4);
        \draw (x4) -- (x1);
        \node  at (1,-3)
        {
            (a) Constraint Graph
        };
        \node at (5,1)  {
           $f(x_1, x_2) = x_{1}^{2} - x_{2}^{2}$
       
        };
        \node at (5,0)  {
             $f(x_1, x_3) = x_{1}^{2} + 2x_{1}x_{3}$
        };
        \node at (5,-1)  {
           $f(x_1, x_4) = 2x_{1}^{2} - 2x_{4}^{2}$
        };
        \node at (5,-2)  {
            $f(x_3, x_4) = x_{3}^{2} + 3x_{4}^{2}$
        };
        \node at (5,2)  {
             $D_i = [-10, 10]$
        };
        \node  at (5.5,-3)
        {
            (b) Cost Functions 
        };
\end{tikzpicture}
\caption{Example of an F-DCOP}
\label{fdcopex}
\end{figure}
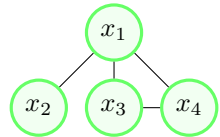
Similar to the DCOP formulation, F-DCOP can be defined as a tuple $ \langle A,X,D,F,\alpha \rangle$. In F-DCOP, $A$, $F$ and $\alpha$ are the same as defined in DCOP. Nonetheless, the set of variables, $X$ and the set of Domains, $D$ are defined as follows - 
\begin{itemize}
    \item X is the set of continuous variables $\{x_1,x_2,...,x_m\}$ that are controlled by agents A.
    \item D is a set of continuous domains $\{D_1, D_2,...,D_m\}$, where each variable $x_i$ can take any value between a range, $D_i$ = [$LB_i$, $UB_i$].
\end{itemize}

As discussed in the previous section, a notable difference between F-DCOP and DCOP is found in the representation of cost function. In DCOP, the cost functions are conventionally represented in tabular form, while in F-DCOP each constraint is represented in the form a function \cite{hoang2019new}. However, the goal remains the same as depicted in Equation \ref{eq:1}. Figure~\ref{fdcopex} presents the example of an F-DCOP where Figure~\ref{fdcopex}a represents the constraint graph with four variables where each of the variable $x_i$ is controlled by an  agent $a_i$.  Each edge in Figure~\ref{fdcopex}a stands for a constraint function and the definition of each function is shown in Figure~\ref{fdcopex}b. Each variable $x_i$ can take values from the range [-10, 10] in this particular example. 

\subsection{Particle Swarm Optimization}
PSO is a population based optimization technique inspired by the movement of a bird flock or a fish school. In PSO,  each individual of the population is called a particle. 
PSO solves the problem by moving the particles in a multi-dimensional search space by adjusting the particle's position and velocity. As shown in Algorithm~\ref{algo:pso}, initially each particle is assigned a random position and velocity. A fitness function is defined which is used to evaluate the position of each particle. For simplicity, we are going to consider the optimization and minimization interchangeably throughout the paper. In each iteration, the movement of a particle is guided by its personal best position found so far in the search space, as well as the global best position found by the entire swarm (Algorithm~\ref{algo:pso}: Lines $4-5$). The combination of the personal best and the global best position ensures that when a better position is found through the search process, the particles will move closer to that position and explore the surrounding search space more thoroughly considering it as a potential solution. The personal best position of each particle and the global best position of the entire population is updated if necessary (Algorithm~\ref{algo:pso}: Lines $6-9$). Over the last couple of decades, several versions of PSO have been developed. The standard PSO has a tendency to converge to a suboptimal solution since the velocity component of the global best particle tends to zero after some iterations. Consequently, the global best position stops moving and the swarm behavior of all other particles leads them to follow the global best particle. To cope with the premature convergence property of standard PSO, Guaranteed Convergence PSO (GCPSO) has been proposed that provides convergence guarantees to local optima \cite{van2002new}. To adapt similar convergence behavior to F-DCOP, we choose to adapt GCPSO in our proposed method.
\begin{algorithm}[t]
\DontPrintSemicolon
\small
Generate an $n$-dimensional population, $P$\;
   Initialize positions and velocities of each particle\;
   \While{Termination condition not met}
   {
   		\For{each particle $P_i \in P$}{
   		    calculate current velocity and position\;
   		    \If{current position $<$ personal best}{
   		    update personal best\;
   		    }
   		    \If{current position $<$ global best}{
   		    update global best\;
   		    }
   		    
   		}
   }
\caption{Particle Swarm Optimization}
\label{algo:pso}
\end{algorithm}
\subsection{Challenges}
The following challenges must be addressed to develop an anytime F-DCOP algorithm that adapts the guaranteed convergence PSO:
\begin{itemize}
\item \textbf{Particles and Fitness Representation:}
We need to define a representation for the particles where each particle represents a solution of the F-DCOP.  Moreover, a distributed method for calculating the fitness for each of the particle need to be devised. 
\item \textbf{Creating Population:} In centralized optimization problems, creating the initial population is a trivial task. But in case of F-DCOP, different agents control different variables. Hence, a method need to be devised to cooperatively generate initial population.
\item \textbf{Evaluation:} Centralized PSO deals with an n-dimensional optimization task. In F-DCOP, each agent holds k variables $(k \leq n)$ and each agent is responsible for solving k-dimensional optimization task where the global objective is still an n-dimensional optimization process. 
\item \textbf{Maintaining Anytime Property:} To maintain the anytime property in a F-DCOP model we need to identify the global best particle and the personal best position for each particle. A distribution method needs to be devised to notify all the agents when a new global best particle or personal best position is found. Finally, a coordination method is needed among the agents to update the position and velocity considering the current best position.
\end{itemize}
In the following section we devise a novel method to apply PSO in F-DCOP while maintaining the balance between local benefit and global benefit.

\section{Proposed Method}
PFD is a PSO based iterative algorithm consisting of three phases: \textit{Initialization}, \textit{Evaluation} and \textit{Update}. In the initialization phase, a pseudo-tree is constructed, initial population is created and parameters are initialized. In the evaluation phase, agents distributedly calculate the fitness function for each particle. The update phase keeps track of the best solution found so far and propagates the information to the agents and updates the assignments accordingly. The detailed algorithm can be found in
Algorithm \ref{algo:PFD}.\par
\begin{figure}[t]

  \begin{tikzpicture}
        [
        roundnode/.style={circle, draw=green!60, fill=green!5, very thick, minimum size=7mm},
        ]
        \node[roundnode]    at(1,0)  (x1)                              {$x_1$};
        \node[roundnode]    at(0,-1)  (x2)                              {$x_2$};
        \node[roundnode]    at(1,-1)  (x3)                              {$x_3$};
        \node[roundnode]    at(2,-1)  (x4)                              {$x_4$};
         
        \draw (x1) -- (x2);
        \draw (x1) -- (x3);
        \draw[dotted] (x3) -- (x4);
        \draw (x4) -- (x1);
        \node  at (1,-2)
        {
            (a) BFS psuedo tree
        };

        \node[roundnode]    at(6,0)  (x1)                              {$x_1$};
        \node[roundnode]    at(5,-1)  (x2)                              {$x_2$};
        \node[roundnode]    at(6,-1)  (x3)                              {$x_3$};
        \node[roundnode]    at(7,-1)  (x4)                              {$x_4$};
                 
        \draw[->] (x1) -- (x2);
        \draw[->] (x1) -- (x3);
        \draw[->] (x3) -- (x4);
        \draw[->] (x1) -- (x4);
        
        \node  at (5.5,-2)
        {
            (b) Ordered arrangement 
        };
\end{tikzpicture}
\caption{Pseudo tree construction and ordered arrangement}
\label{bfsptree}
\end{figure}
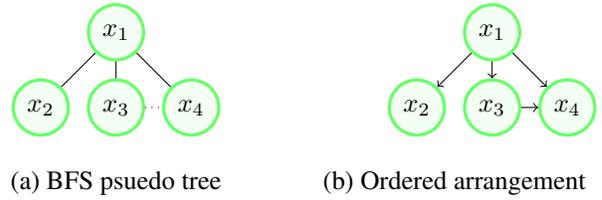
\textbf{Initialization} starts with ordering the agents in a Breadth First Search(BFS) pseudo-tree \cite{chen2017improved}. The pseudo-tree serves the purpose of defining a message passing order which is used in the evaluation and update phase. In the ordered arrangement the agents with lower depths have higher priorities over the agents with higher depths and ties are broken in the alphabetical order. Figure \ref{bfsptree}(a) and (b) illustrates the BFS pseudo tree and the ordered arrangement. In Figure \ref{bfsptree}(b), $x_1$ is the root and the arrows represent the message passing direction. From this point for an agent $a_i$, we refer $N_i$ as the set of neighbors, $H_i \subseteq N_i$ and $L_i \subseteq N_i$ as the set of higher priority and lower priority neighbors respectively. In Figure \ref{bfsptree}(b) for agent $x_3$, $N_i = \{x_1, x_4\} $, $H_i = \{x_1\}$ and $L_i = \{x_4\}$.\par
At the beginning of algorithm, PFD takes input from the users to initialize all the parameters where $K$ is the number of particles. The parameters depend on the experiments; the recommended settings for our experiments are discussed later in the text. We define $P$ as the set of K particles which is maintained by the agents where each agent holds component(s) of the particles. Each particle $P_k \in P$ has a velocity and a position attribute. The velocity attribute defines the movement directions and position attribute defines the variable values associated with the variables that the agent holds.
Then each agent $a_i$ executes \textbf{Init}(Algorithm \ref{algo:PFD}: Lines \ref{alg:Finit1} to \ref{alg:Finit2}) and initializes the the velocity component, $v_i$ to $0$ and position component, $x_i$ to a random value from its domain $D_i$ for each particle $P_k$.  For the example of Figure \ref{bfsptree}(b), let us assume number of particles, $K=2$,  and the set of particles, $P = \{P_1, P_2\}$ . Here, $P_1.V = P_2.V = \{0, 0, 0, 0\}$ shows the complete assignment for velocity attribute of two particles and the complete assignment for position attribute can be shown as, $P_1.X= \{x_1 = -1, x_2 = 0, x_3 = 2, x_4 = 9.5\},  P_2.X = \{x_1 = 3.5, x_2 = 4.9, x_3 = 1, x_4 = 0\}$. We define $P_k.x_i$ and $P_k.v_i$ as the position and velocity component of particle $P_k$ set by agent $a_i$. In this example $P_1.x_3 = 2$ which is the value of variable $x_3$ of particle $P_1$ set by agent $a_3$. After selecting the value of its variable each agent shares the particle set, $P.x_i$ to its lower priority neighbors, $L_i$. For this example, agent $a_3$ sends $P.x_3 = \{2, 1\}$ to its lower priority neighbor $a_4$.\par
\begin{algorithm}[!h]
\DontPrintSemicolon
\small
\SetKwFunction{FUpdate}{Update}
\SetKwFunction{FInit}{Init}
   Construct BFS pseudo-tree\;
   Initialize parameters: $K, w, c_1, c_2, max_{s_{c}}, max_{f_{c}}$\;
   $P \leftarrow$ set of $K$ particles\;
   \SetKwProg{In}{Function}{:}{} \label{alg:Finit1}
  \In{\FInit{}}{
        \For{each particle $P_k \in P$}{
   		    $P_k.v_i \leftarrow$ 0\;
   		    $P_k.x_i \leftarrow$ a random value from $D_i$ \;
   		}
   		Sends $P.x_i$ to agents in $L_i$\;
  } \label{alg:Finit2}
   \For{each agent $a_i$}{
   		\FInit{}\;
   	}
   \While{Termination condition not met each agent $a_i$}
   {
        
   	    \For{$P.x_i$ received from $H_{i{_j}} \in H_i$ }{
   	         \For{each particle $P_k \in P$ }{
					$P_k.fitness \leftarrow Cost_{i,j}(P_k.x_i,P_k.x_j) $ \;
             }
             Sends $P.fitness$ to agents in $H_{i{_j}}$\;
             
        }
       
       Wait until $P.fitness$ received from all agent in $L_i$\;
        \If{$|L_i| \not=0$ and $P.fitness$ received from all agent in $L_i$}{   	        
   	         \For{each particle $P_k \in P$ }{
                $P_k.fitness \leftarrow \sum_{j\in L_i}^{} P.fitness $\;
                
            }
            \If{$a_i \not= $ root}{
           Sends $P.fitness$ to an $H_{i{_j}} \in H_i$\;
        }
            }
            
             \If{$a_i =$ root} {
            \FUpdate{$P.fitness$}\;
        }
   	    
      Wait until $P.p_{best}$ and $P.g_{best}$ receives from  $H_i$\;
     \If{$P.p_{best}$ and $P.g_{best}$ receives from  $H_i$}{
     Calculate $s_c$ and $f_c$ according to equation \ref{eq:7}, \ref{eq:8}\;
        \For{each particle $P_k \in P$}{
            \eIf{$P.g_{best}=P_k$}{
                Calculate $P_k.v_i$ and $P_k.x_i$ according to equation \ref{eq:3}, \ref{eq:5}\;
            }
            {
            Calculate $P_k.v_i$ and $P_k.x_i$ according to equation \ref{eq:4}, \ref{eq:5}\;
            }
            
        }
        
        \If{$|L_i| \not = 0$}{
        Sends $P.x_i$ to agents in $L_i$\;
            Sends $P.p_{best}$ and $P.g_{best}$ to agents in $L_i$\;
        }
     }
   }
  \SetKwProg{Fn}{Function}{:}{}
  \Fn{\FUpdate{$P.fitness$}}{
        $P.p_{best} \leftarrow \{\}$
        
        \For{each particle $P_k \in P$ }{
                \If{$P_k.fitness < P_k.p_{best}.fitness$}{
                $P_k.p_{best} \leftarrow P_k$\;
                $P.p_{best} \leftarrow P_k.p_{best} \cup P.p_{best}$\;
                }
                \If{$P_k.fitness < P.g_{best}.fitness$}{
                $P.g_{best} \leftarrow P_k$\;
                }
            }
            Sends $P.p_{best}$ and $P.g_{best}$ to agents in $L_i$\;
  }
\caption{Particle Swarm F-DCOP}
\label{algo:PFD}
\end{algorithm}
\vspace{-1mm}
\textbf{Evaluation} phase of PFD calculates the fitness of each particle, $P_k$ using a fitness function shown in Equation \ref{eq:2} where $P_k.X$ represents the complete assignment of variables in $X$. 
\begin{equation}
    P_k.fitness = \sum_{f_i\in F}^{} f_i(P_k.x^i) 
\label{eq:2}
\end{equation}
This phase starts after the agents receive value assignments from all the higher priority neighbors. Each agent $a_i$ is responsible for calculating the constraint cost associated with each of its higher priority neighbors from $H_i$ (Algorithm \ref{algo:PFD}: Lines 13-16). We define $P_k.fitness$ as the local fitness of each particle $P_k$ of the particle set $P$. When an agent $a_i$ receives value assignments $P.x_i$, from a higher priority neighbor $H_{i_{j}} \in H_i$, it calculates the constraint cost between them and sends it to $H_{i{_j}}$. Additionally, each agent except the leaf agents need to pass the constraints cost upward the pseudo-tree calculated by lower priority neighbors, $L_i$ (Algorithm \ref{algo:PFD}: Lines 18-19)\par
For the example shown in Figure \ref{fdcopex}, agent $a_4$ sends the fitness $\{P_1 = 274.75, P_2 = 1\}$ to $a_3$ and fitness $\{P_1 = -178.5, P_2 = 24.5\}$ to $a_1$. Agent $a_2$ calculates the fitness $\{P_1 = -3, P_2 = 19.25\}$ and sends it to $a_1$. Furthermore, $a_3$ receives the fitness from $a_4$ and passes it to $a_1$. Similarly, $a_2$ sends the fitness $\{P_1 = 1, P_2 = -11.76\}$ to $a_1$.\par
\textbf{Update} phase consists of two parts: $p_{best}$, $g_{best}$ update and variable update. We define $p_{best}$ to be the personal best position achieved so far by each particle and $g_{best}$ to be the global best position among all the particles. Since each agent calculates and passes the cost of the constraints to the agents in $H_i$, the fitness of all the particles propagate to the root. The root agent then sums the fitness values received from the agents in $L_i$ for each of the particles, $P_k$. Then the root agent checks and updates the $p_{best}$ for $P_k \in P$ and $g_{best}$ for $P$ and sends the new values to the agents in $L_i$ (Algorithm \ref{algo:PFD}: Lines 38-44). 
When an agent $a_i$ receives $p_{best}$ and $g_{best}$ of the previous iteration, it updates the the velocity component $P_k.v_i$ and position component $P_k.x_i$ for $P_k \in P$. To adapt the guaranteed convergence method to PFD, two types of update equations for velocity component are defined. If the particle is the current global best particle, the update equation is defined as follows:
\begin{equation}
\begin{split}
P_k.v_i(t) = -P_k.x_i(t-1) + P.g_{best}(t-1) + \\ w * P_k.v_i(t-1) + \rho*(1-2r_2)
\end{split}
\label{eq:3}
\end{equation}
For all other particles, the velocity update equation is defined as follows:
\begin{equation}
\begin{split}
P_k.v_i(t) = w * P_k.v_i(t-1) + r_1 * c_1 * \\ (P_k.p_{best}(t-1) - P_k.x_i(t-1)) + r_2 * c_2 *\\ (P.g_{best}(t-1) - P_k.x_i(t-1))
\end{split}
\label{eq:4}
\end{equation}
The position component update equation is same for all the particles which is defined in the following equation:
\begin{equation}
    P_k.x_i(t) = P_k.x_i(t-1) + P_k.v_i(t)
\label{eq:5}
\end{equation}
In equations \ref{eq:3} \ref{eq:4}, and \ref{eq:5}, $P_k.v_i(t)$ and $P_k.x_i(t)$ refers to the velocity and position components controlled by agent $a_i$ for particle $P_k$ in $t^{th}$ iteration. Here, an iteration refers to a complete round of the Evaluation and Update phase (Algorithm \ref{algo:PFD}: Line 12). $w$ is the inertia weight which defines the influence of current velocity on the updated velocity, $r_1$ and $r_2$ are two random values between [0, 1] and $c_1$, $c_2$ are two constants. Combinations of $c_1$ and $c_2$ define the magnitude of influence personal best and global best have on the updated particle position. In equation \ref{eq:3}, $\rho$ is used to explore a random area near the position of the global best particle. To be precise, $\rho$ defines the diameter of this area that the particles can explore. The value of $\rho$ is adjusted according to equation \ref{eq:6}.
\begin{equation}
\rho(t) = 
 \begin{cases} 
      1 & t=0\\
      2*\rho(t-1) & s_c > max_{s_{c}} \\
      0.5*\rho(t-1) & f_c > max_{f_{c}} \\
      \rho(t-1) & otherwise 
   \end{cases}
\label{eq:6}
\end{equation}
\par
In equation \ref{eq:6}, $s_c$ and $f_c$ are the count of consecutive success and failures respectively. A success is defined when the global best particle updates its personal best position. Similarly, a failure is defined when the position of the global best particle remains unchanged. The parameters $max_{s_{c}}$ and $max_{f_{c}}$ are the upper bound of $s_c$ and $f_c$. The following equations define $s_c$ and $f_c$.\par
\begin{equation}
    s_c(t) = 
    \begin{cases} 
      s_c(t-1) + 1 & if \ P_G.p_{best}(t) <P.g_{best}(t-1)  \\
      0 & otherwise 
  \end{cases}
\label{eq:7}
\end{equation}
\begin{equation}
    f_c(t) = 
    \begin{cases} 
      f_c(t-1) + 1 & if \ P.g_{best}(t) = P.g_{best}(t-1)  \\
      0 & otherwise 
  \end{cases}
\label{eq:8}
\end{equation}
In equation \ref{eq:7}, $P_G$ defines the global best particle of iteration $t-1$. Each agent calculates $s_c$ and $f_c$ according to equations \ref{eq:7} and \ref{eq:8} after receiving $P_G.p_{best}$ and $P.g_{best}$ from their higher priority neighbors, $H_i$ (Algorithm \ref{algo:PFD}: Line 27). \par
Consider agent $a_1$ in Figure \ref{bfsptree}. When $a_1$ receives fitness value from all of its lower priority neighbors, it is ready to calculate the $P.p_{best}$ and $P.g_{best}$. The final updated fitness value, $P.fitness = \{94.25, 33\}$. Based on the updated values $a_i$ constructs $P.p_{best} = \{94.25, 33\}$ and $P.g_{best}$ = 33 and notifies the agents in $L_i$. Then each agent calculates $s_c$ and $f_c$ and updates the values based on equation \ref{eq:3}, \ref{eq:4}, and \ref{eq:5}.

\section{Theoretical Analysis}
In this section, we first prove PFD is an anytime algorithm that is, solution quality improves and never degrades over time. Later, we provide the theoretical complexity analysis in terms of communication, computation and memory.\par
\textbf{Lemma 1:} At iteration \footnote{For the theoretical analysis section, iteration refers to the communication steps required. In one communication step agents only directly communicate with the neighbors.} $t+d$ , root is aware of the $P.p_{best}$ and $P.g_{best}$ up to iteration $t$, where $d$ is the longest path between root and any node in the pseudo-tree.\par
To prove this lemma it is sufficient to show that, at iteration $t+d$, root agent has enough information to calculate $P.p_{best}$ and $P.g_{best}$ up to iteration $t$, that is, root agent knows the fitness of each particle. To calculate the fitness of each particle $P_k.fitness$ according to equation \ref{eq:2}, the root agent needs cost messages from the agents in $L_{a_{i}}$. The cost messages from agents at $d$ distance from root will need $d-1$ iteration to reach agents in $L_{a_{i}}$. By induction, it will take $t+d$ iterations to reach the cost messages calculated at iteration $t$ from the agents with distance $d$ to root.\par
\textbf{Lemma 2:} At iteration $t+2d$, each agent is aware of the $P.p_{best}$ and $P.g_{best}$ up to iteration $t$\par
In PFD, for any agent $a_i$, the value message passing length and cost message passing length from the root are same. So, it takes $d$ iterations to reach the $P.p_{best}$ and $P.g_{best}$ to the agents at distance $d$ from the root. Using lemma 1 and the above claim, it takes $t+d+d = t+2d$ iterations to reach $P.p_{best}$ and $P.g_{best}$ to the agent at $d$ distance from the root.\par
\textbf{Proposition 1:} PFD is an anytime algorithm.\par
By lemma 2, at iteration $t+2d$ and $t+2d+\delta$ ($\delta \geq 0)$ each agent is aware of the $P.p_{best}$ and $P.g_{best}$ up to iteration $t$ and $t+\delta$ respectively. Let us assume, $P.p_{best}$ and $P.g_{best}$ at iteration  $t+2d+\delta \geq t+2d$. But for any $\delta \geq 0$, $t+2d+\delta >= t+2d$ and $P.p_{best}$ and $P.g_{best}$ only gets updated if a better solution is found. Therefore, using proof by contradiction, $P.p_{best}$ and $P.g_{best}$ at iteration $t+2d+\delta \leq t+2d$ that is, solution quality improves monotonically as the number of iterations increases. Thus we prove, PFD is an anytime algorithm.

\subsection{Complexity Analysis}
We define, the total number of agents $|a| = n$ and the total number of neighbors of an agent $a_i \in a$, $|N_i| = |L_i| + |H_i|$. In PFD, during the Initialization and Update phase an agent sends $|L_i|$ messages. Additionally, during the Evaluation phase an agent sends $|H_i|+1$ messages. After one round of completion of Initialization, Evaluation and Update phases, an agent $a_i$ sends $2*|L_i| + |H_i| = |L_i| + |N_i| $ messages. In the worst case, the graph is complete where $|N_i| = n$. In a complete graph if $|L_i| = n$, then $|H_i| = 0$. Therefore, the total number of messages sent by an agent $a_i$ is $O(2*|L_i| + |H_i|) = O(2n)$ in the worst case.\par
The size of each message can be calculated as the size of each particle multiplied by the number of particles. If the total number of particle is $K$, at each iteration the total message size for an agent $a_i$ is $O(K*n*2n) = O(n^2)$ in the worst case.\par
During an iteration, an agent only needs to calculate $P_k.v_i$ and $P_k.x_i$ for each of the particle $P_k$. Hence, the total computation complexity per agent during an iteration is $O(2*K) = O(K)$ where K is the number of particles.

\section{Experimental Results}
In this section, we empirically evaluate the quality of solutions produced by PFD with HCMS and AF-DPOP on two types of graphs: \textit{Random Graphs} and \textit{Random Trees}. However, 
CMS is not used in comparison because it only works with peicewise linear functions which is not applicable in most of the real world applications. Although Hoang et al., proposed three versions Functional DPOP, we only compare with AF-DPOP here. The reason is AF-DPOP is reported to provide the best solution among the approximate algorithms proposed in their work. 
For the experimental performance evaluation, binary quadratic functions are used which are of form $ax^2 + bxy + cy^2$. Note that, although we choose binary quadratic functions for evaluation, PFD is broadly applicable to other class of problems. The experiments are carried out on a laptop with an Intel Core i5-6200U CPU, 2.3 GHz processor and 8 GB RAM.
The detailed experimental settings are described below.\par
\begin{figure}[t]
\centering
  \includegraphics[scale = 0.60]{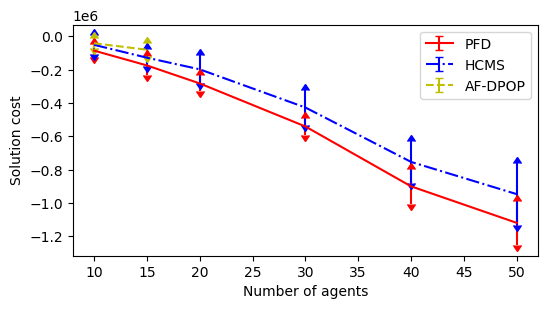}
  \vspace{-3mm}
  \caption{Solution Cost Comparison of PFD and the competing algorithms varying number of agents (sparse graphs)}
  \label{fig:sparse}
  \vspace{2.5mm}
\end{figure}

\begin{figure}[t]
\centering
  \includegraphics[scale = 0.60]{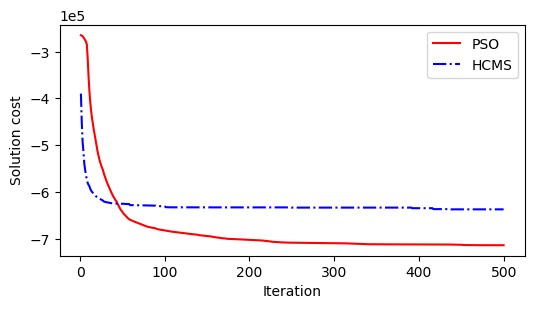}
  \vspace{-3mm}
  \caption{Solution Cost Comparison of PFD and the competing algorithms with iterations (sparse graphs)}
  \label{fig:iter}
  \vspace{2.5mm}
\end{figure}
\begin{figure}[t]
\centering
  \includegraphics[scale = 0.60]{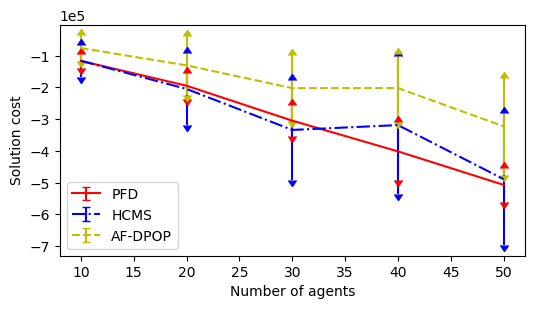}
  \vspace{-3mm}
  \caption{Solution Cost Comparison of PFD and the competing algorithms varying number of agents (scale-free graphs)}
  \label{fig:scalefree}
  \vspace{2.5mm}
\end{figure}
\begin{figure}[t]
\centering
  \includegraphics[scale = 0.60]{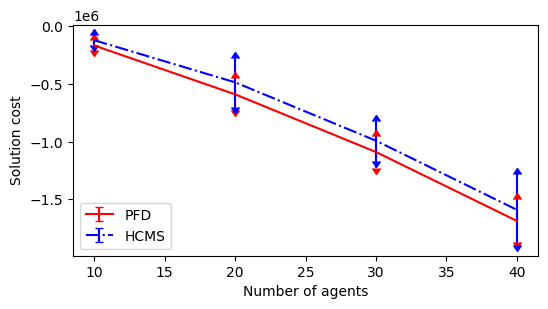}
  \vspace{-3mm}
  \caption{Solution Cost Comparison of PFD and the competing algorithms varying number of agents (dense graphs)}
  \label{fig:dense}
  \vspace{2.5mm}
\end{figure}
\textbf{Random Graphs:} For random graphs we use three settings - sparse, dense and scale-free. Figure \ref{fig:sparse} shows the comparison of average costs on Erd{\H{o}}s-R{\'e}nyi topology with sparse settings (edge probability 0.2) varying the number of agents. We choose coefficients of the cost functions $(a, b, c)$ randomly between $[-5, 5]$ and  set the domains of each agents to $[-50, 50]$. For our proposed algorithm PFD, we set the parameters, $K = 2000$, $w = 0.9$, $c_1 = 0.9$, $c_2 = 0.1, max_{f_{c}} = 5,$ and $max_{s_{c}} = 15$. For both HCMS and AF-DPOP we choose the number of discrete points to be 3. The discrete points are chosen randomly between the domain range. The averages are taken over 50 randomly generated problems. Figure \ref{fig:sparse} shows that PFD performs better than both HCMS and AF-DPOP on average. Notably, the performance of HCMS varies significantly which results in a high standard deviation. The reason behind the high standard deviation is that, the performance of HCMS on cyclic graph varies on the initial discretization of domains of the agents. For $no. \ of \ agents \geq 20$, AF-DPOP ran out of memory. Thus, we omit the result of AF-DPOP for $no. \ of \ agents \geq 20$.\par

Figure \ref{fig:iter} shows the comparison between PFD and HCMS on sparse graph settings with increasing number of iterations. We set the number of agents to 50 and other settings are same as the above experiment. Moreover, we stop both algorithms after 500 iterations. HCMS initially performs slightly better than PFD till 50 iterations since the particles of PFD initially start from random positions and require few iterations to move the particles towards the best position. However, PFD outperforms HCMS later and the improvement rate of PFD is steadier than HCMS. Note that, for 50 agents, AF-DPOP run out of memory in our settings. Thus, we omit the result of AF-DPOP here.\par
To compare with the performance of AF-DPOP on larger graphs we use scale-free graphs. Figure \ref{fig:scalefree} shows the average cost comparison between the three algorithms with increasing number of agents. PFD shows a comparable performance with HCMS upto 30 agents and outperforms HCMS afterwards. Both PFD and HCMS outperforms AF-DPOP. The huge standard deviation of HCMS results into the comparable performance with PFD for smaller agents.\par
We choose dense graphs as our final random graph settings. Figure \ref{fig:dense} shows comparison between the PFD and HCMS on Erd{\H{o}}s-R{\'e}nyi topology with dense settings (edge probability 0.6). PFD shows comparatively better performance than HCMS. Note than, AF-DPOP is not used in dense graph due to the huge computation overhead.\par
\begin{figure}[t]
\centering
  \includegraphics[scale = 0.60]{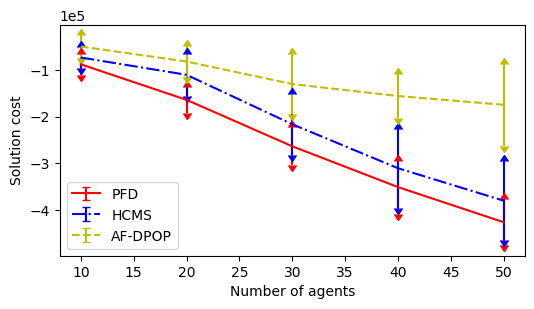}
  \vspace{-3mm}
  \caption{Solution Cost Comparison of PFD and the competing algorithms varying number of agents (random trees)}
  \label{fig:tree}
  \vspace{2.5mm}
\end{figure}

\textbf{Random Trees:} We use the random tree configuration in our last experimental settings since the memory requirement of AF-DPOP is less on trees. The experimental configurations are similar to the random graph settings. Figure \ref{fig:tree} shows the comparison graph between PFD and the competing algorithms on random trees. The closest competitor of PFD in this setting is HCMS. On an average, PFD outperforms HCMS which in turn outperforms AF-DPOP. When the number of agent is 50, PFD shows better performance than AF-DPOP at a significant level. 
\section{Conclusions}
In order to model many real world problems, continuous valued variables are more suitable than discrete valued variables. F-DCOP framework is a variant of DCOP framework that can model such problems effectively. To solve F-DCOPs, we propose an anytime algorithm called PFD that is inspired by Particle Swarm Optimization (PSO) technique. 
To be precise, PFD devises a new method to calculate and propagate the best particle information across all the agents which influence the swarm to move towards a better solution. We also theoretically prove that our proposed algorithm PFD is anytime. Moreover, the guaranteed convergence version of PSO is  tailored in PFD  which ensures its convergence to a local optima. We empirically evaluate our algorithm in a number of  settings, and compare the results with the state-of-the-art algorithms, HCMS and AF-DPOP. In all of the settings, PFD markedly outperforms its counterparts in terms of solution quality. In the future, we would like to further investigate the potential of PFD on various F-DCOP applications. We also want to explore whether PFD can be extended for multi-objective F-DCOP settings.  
\clearpage
\bibliographystyle{aaai}
\bibliography{pso_dcop}
\end{document}